\documentclass[11pt]{amsart}

\usepackage{hyperref}
\usepackage{amssymb,amsfonts,latexsym,amsmath,amsthm,wrapfig,graphicx}
\input xy
\xyoption{all}

\def\la{\label}
\def\be{\begin{equation}}
\def\beq{\begin{equation}}
\def\eeq{\end{equation}}
\def\ee{\end{equation}}
\def\bea{\begin{eqnarray}}
\def\eea{\end{eqnarray}}

\begin{document}

\title[Atmospheric carbon dioxide: Europe study]{Contributors of carbon dioxide in the atmosphere in Europe}

\author{Iuliana Teodorescu}
\author{Chris Tsokos}
\address{Statistics Department, University of South Florida, Tampa Florida}

\begin{abstract}
Carbon dioxide, along with atmospheric temperature are interacting to cause what we have defined as {\bf{global warming}}. In the present study we develop a statistical model using real data to identify the attributable variables (risk factors) that cause the CO$_2$ emissions in the atmosphere in Europe. Some scientists believe that there are more than nineteen attributable variables that cause the CO$_2$ in our atmosphere. However, our study has identified only three individual risk factors and five interactions among the attributable variables that cause almost all the CO$_2$  emissions in the atmosphere in Europe. We rank the risk factors and interactions according to the amount of CO$_2$  they generate. In addition, we compare the present findings of the Eupean data with a similar study for the Continental United States \cite{model, article1}. For example, in the US, liquid fuels ranks number one, while in Europe is gas fuels. In fact, liquid fuels in Europe is the least contributable variable of CO$_2$ in the atmosphere, and gas fuels ranks seventh.

%Starting from the atmospheric CO$_2$ measurements taken in Hawaii between 1959 and 2008, a quadratic model with interactions was fitted for the attributable variables recorded in a group of 23 countries that are members of the European Union. Surface response analysis returned the eigenvalues and eigenvectors at the critical point, which turns out to be of mixed type, with two positive eigenvalues and two negative. From these data, it is derived that the relevant attributable variables and interactions are not the same for Europe and USA, and that the surface response analysis leads to a very different picture regarding the environmental impact of various attributable variables and their combinations.
\end{abstract}
%\date{March 2013}
\maketitle

%\tableofcontents
%\listoffigures
%\listoftables

\section{Introduction} \label{sec:intro} %\markright{Introduction}

The proposed model that we are developing takes into consideration individual contributions and interactions along with higher order contributions if applicable. In developing the statistical model, the response variable is the CO$_2$ in the atmosphere and is given in parts per million by volume (ppmv). In the present analysis, we used real yearly data that has been collected from 1959 to 2008\footnote{Year 1964 was ignored due to incomplete records.} at Mauna Loa Observatory, Hawaii. The CO$_2$ emission data for the EU countries (list given below) was obtained from Carbon Dioxide Information Analysis Center (CDIAC) during the same period. 
The analysis presented here consists of two steps: first, we partially replicate the comprehensive study performed in \cite{model, article1} in order to select the relevant variables and their interactions, and then we validate and analyze the best second-order model. Based on this result, we will perform the surface-response analysis (curvature effects) of the model in a forthcoming publication \cite{article3}. 

\begin{figure}%[h!!!!!]
\begin{center}
 \includegraphics*[width=12cm]     {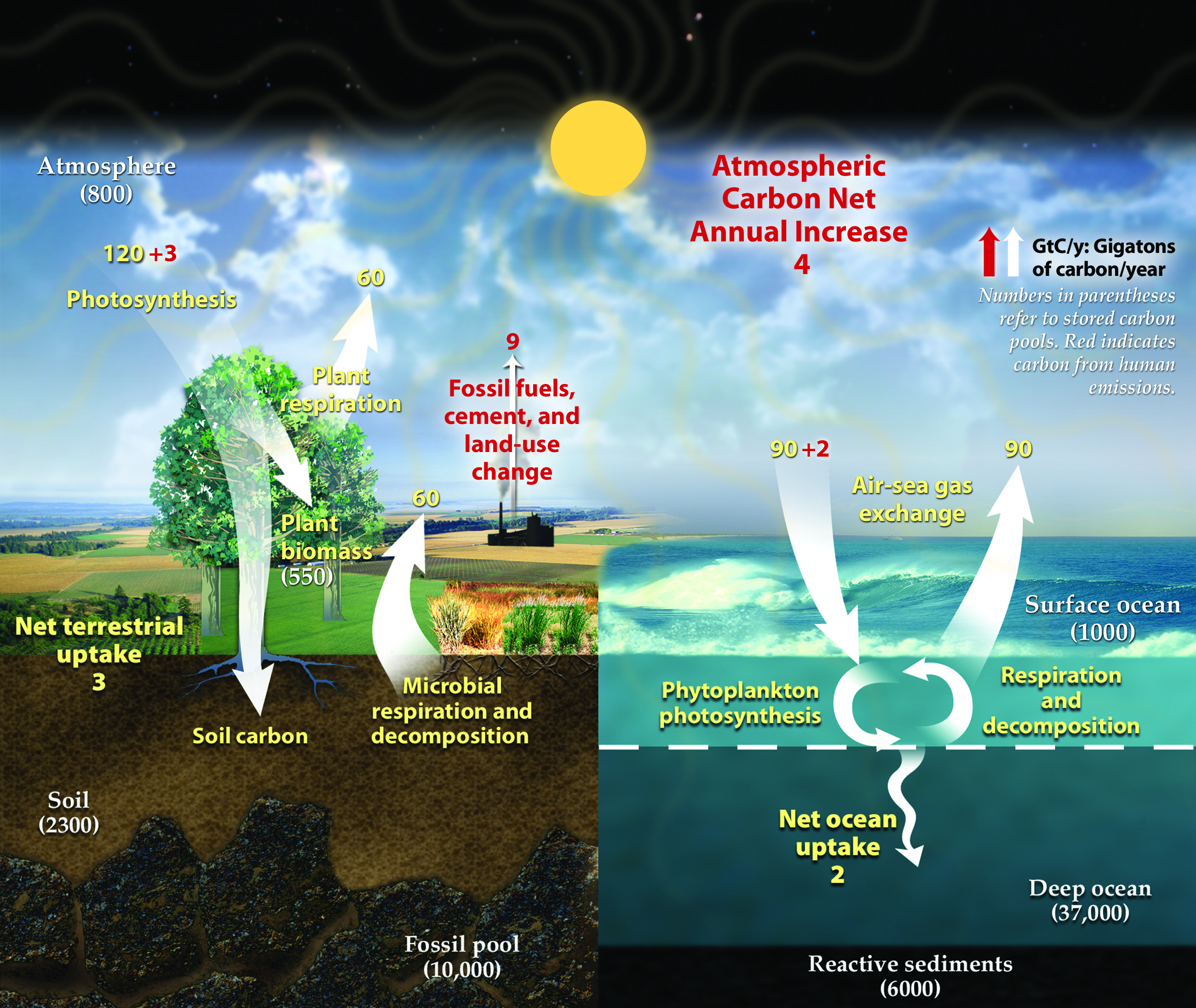}
\caption{Schematic representation of the carbon cycle (DOE report \cite{ipcc} and supporting documentation).}
\label{ip}
\end{center} 
\end{figure}

\begin{figure}[h!!!!!]
\begin{center}
 \includegraphics*[width=12.5cm]     {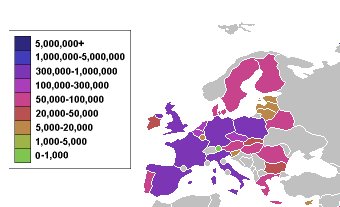}
\caption{EU CO$_2$ emissions, in thousands of metric tons \cite{pic}.} 
\label{map}
\end{center} 
\end{figure}

The following 23 EU countries\footnote{As of June 2013, there are 27 member states of the EU. Slovenia and the Baltic states were excluded from the present study since there was no individual data available for the period during which they were part of former Yugoslavia, and former Soviet Union, respectively. However, their contribution to the CO$_2$ emissions is relatively small, as it can be seen from Figure~\ref{map}.} were used in this study\footnote{Figure~\ref{map} was reproduced under the  Creative Commons Attribution-Share Alike 3.0 Unported license.}: 
Austria,
Belgium,
Bulgaria,
Cyprus,
Czech Republic,
Denmark,
Finland,
France,
Germany,
Greece,
Hungary,
Ireland,
Italy,
Luxembourg,
Malta,
Netherlands,
Poland,
Portugal,
Romania,
Slovakia,
Spain,
Sweden,
United Kingdom.

The aim of this study is two-fold: in the current paper we focus on the determination of the optimal second-order model 
(including interactions as well as quadratic terms) which provides the best fit for the data and features robust validation; in a forthcoming publication, we will use the model derived here in order to determine multi-variables confidence regions and obtain estimates for the allowed fluctuations in attributable variables, given a maximal range of change in the response variable.  The second-order model we will obtain in the current study provides us with a list of relevant variables, ranked according to their statistical significance (quantified by the respective contribution to the total variability), which we employ in order to compare with the similar result obtained in the case of the U.S. data \cite{model, article1}.

\section{Regression analysis and model building}

One of the underlying assumptions to construct the model is that the response variable should follow Gaussian distribution. It is known \cite{model, article1} that the CO$_2$ in the atmosphere does not follow the Gaussian distribution. 
Therefore, the Box-Cox transformation is applied to the CO$_2$ atmosphere data to filter the data to be normally distributed. After the Box-Cox filter, we retest the data and it shows our data will follow normal distribution; thus, we proceed to estimate the coefficients of the contributable variables for the transformed CO$_2$ atmosphere data. The parameter of the transformation is the same as the one used in the two previous studies \cite{model, article1}. 

We can proceed to estimate the approximate coefficients of the contributable variables for transformed CO$_2$ in the atmosphere and obtain the coefficients of all possible interactions, using the multivariate regression procedure and corresponding goodness-of-fit measures.

%
%\begin{figure}%[h!!!!!]
%\begin{center}
%% \includegraphics*[width=9.5cm]     {co_circulation.jpg}
% \includegraphics*[width=6.25cm]     {data_co2.jpg}
% \includegraphics*[width=6.25cm]     {qq_co2.jpg}
%\caption{Atmospheric CO$_2$ data taken at the Mauna Loa station (first panel) and its corresponding $Q-Q$ plot (second panel). Data obtained from CDIAC-ORNL.}
%\label{qq}
%\end{center} 
%\end{figure}

%At the same time, we can determine the significant contributions of both attributable variables and interactions.
We begin with seven attributable variables and apply the SAS stepwise forward selection procedure to a model with six relevant variables and 21 second-order interactions between each pair and self-interactions. Introducing the notation  $x_1 = $ Liquid Fuels (Li), $x_2 = $ Gas Fuels (Ga), $x_3 = $ Gas Flares (Fl), $x_4 = $ Bunker (Bu) for the relevant attributable variables, we find that only three of the variables (Liquid Fuels ($x_1$), Gas Fuels  ($x_2$), and Gas Flares  ($x_1$)) and only 3 interactions (Ga:Bu, Li:Fl,  Li:Bu) and two quadratic terms (Li$^2$, Bu$^2$) are statistically relevant at $\alpha = 0.01$ level.

The result of estimation becomes the quadratic model with interactions:
%(fully consistent with the results of \cite{model}):

\begin{eqnarray*} 
[\widehat{CO}_2]^{-2.376}&  = & 0.00000123 + (710.85 Fl -30.64Ga -3.4501 Li )  \times 10^{-13} + \\
& + & (37.34Ga\cdot Bu + 1.35 Li \cdot Li -65.12 Bu \cdot Bu - \\
& - &  133.05 Li \cdot Fl -5.35 Li \cdot Bu)\times 10^{-18}.
\end{eqnarray*}

The quality of the fit for this quadratic model is evidenced by high value of both $R^2$ and $R^2_{adjusted}$ which are the key criteria to evaluate the model fitting. In terms of the total ($SS_t$), regression ($SS_r$ and error ($SS_e$) sums of squares, we have the standard formulas
$$
R^2 = \frac{SS_r}{SS_t}, \quad R^2_{adjusted} = 1 - \frac{SS_e/df_e}{SS_t/df_t},
$$ 
with $df_{e, t}$ the degrees of freedom for the chi-squared distributions for error and total, respectively \cite{west}. We also employ the prediction of residual error sum of squares ($PRESS$) statistics which will evaluate how good the estimation will be if each time we remove one data point. %If the diagonal entries of the hat matrix are denoted by $H_{ii}$, where 
If the index $i$ covers all the observations and $\hat{y}_{(i)}$ is the predicted value when the observation is omitted, then \cite{allen}
$$
PRESS = \sum_{i}\left ( y_i-\hat{y}_{(i)}\right )^2.
$$

 \begin{center}
 \begin{table}[h!]
  \caption{Statistical evaluation criteria for model \eqref{mex}}
 \begin{tabular}{| c | c | c | }
 \hline 
 $R^2$ & $R^2_{{ adjusted}}$ & ${PRESS}$ \\
 \hline 
\hline 
$   0.9979 $  & $ 0.9975 $ & $1.636 \times 10^{-20}$    \\
 \hline 
 \end{tabular}
 \end{table}
\end{center}

According to these goodness-of-fit measures, the model we have obtained is high quality and reliable for predictions. The model becomes:

\be \label{mex}
[\widehat{CO}_2]^{-2.376} = \beta_0 + \sum_{i=1}^4 \beta_i x_i + \sum_{i \le j = 1}^4 \beta_{ij}x_i x_j, 
\ee
%where the measure for goodness-of-fit ($R^2 = 0.9965$ and the $p-$value less than 0.0001), as well as parameters $\{\beta_k \}$, are found from the  SAS output; here,
with  the corresponding ranks determined by the stepwise  SAS procedure are given in Table~\ref{coeff}, along with the coefficients in the final regression model.

Therefore, we can write our model in matrix notation  (where prime denotes transposition) as: 

\be \la{model}
Y = \beta_0 + \beta' \cdot X + X' \cdot B \cdot X,
\ee
with the obvious identifications 

$$X' = (x_1, \ldots, x_4),  \,\, \beta' = (\beta_1, \ldots, \beta_4), \,\, B_{ij} = B_{ji} = \frac{1}{2}\beta_{ij}\,\, (i < j).$$

 \begin{center}
 \begin{table}[h!] \label{coeff}
  \caption{Ranking by statistical relevance for attributable variables and interactions.}
 \begin{tabular}{| c || l | c | c | c |}
 \hline 
${\mbox{Rank}}$ & ${\mbox{Variable}} $ & Name &  $\beta [\times 10^{-18}]$ & Variation  \\
 \hline 
\hline 
$  1 $  & Ga ($x_ 2$) & Gas Fuels & $-30.635 \times 10^{5}$  & $ 48.72\% $ \\
\hline 
$  2 $  & Ga:Bu ($x_ 2 \times x_4$) & Gas $\times$ Bunker & $37.3391$  & $12. 41\%$   \\
\hline 
$  3 $  & Li$^2$ ($x_ 1^2$) & Liquid $\times$ Liquid & $1.35565$  & $11.79\% $   \\
\hline 
$  4 $  & Bu$^2$ ($x_ 4^2$) & Bunker & $-65.115$  & $7.78\%$   \\
\hline 
$  5 $  & Fl ($x_ 3$) & Gas Flares & $710.848 \times 10^{5}$  & $6.66\% $   \\
\hline 
$  6 $  & Li:Fl ($x_1 \times x_3 $) & Liquid $\times$ Flares & $-133.05$  & $5.06\% $  \\
\hline 
$  7 $  & Li:Bu ($x_1 \times x_4 $) &Liquid $\times$ Bunker & $-5.3501$  &   $4.71\%$ \\
 \hline 
$  8 $  & Li ($x_1$) & Liquid Fuels & $-3.4501 \times 10^{5}$  & $2.86\%$ \\
 \hline 
 \end{tabular}
 \end{table}
\end{center}

More precisely, the vector $\beta$ (up to an overall scale factor of $10^{-13}$), and the  symmetric matrix $B$ (up to an overall scale factor of $10^{-18}$) have the forms:

$$
\beta = 
\left [ 
\begin{array}{c}
-3.4501 \\  -30.635 \\ 710.848 \\ 0
\end{array}
\right ],
\, 
B = 
\left [ 
\begin{array}{cccccc}
2.7113 & 0 &  -133.05 & 0 \\
0 & 0 &  0 & 37.3391 \\
-133.05 & 0 &  0 & 0 \\
0 & 37.3391 & 0 & -130.23 \\
\end{array}
\right ]
$$

These results will be used to perform the surface response analysis for the model, in a separate publication. 
The matrix formulation of the quadratic expression \eqref{mex} will be the starting point for  finding its canonical decomposition, based on the eigenvalue analysis \cite{article3}. 

\subsection{Validation of the fitted model}

%Residual plots for \eqref{mex} are shown in Figure~\ref{res}.

\begin{figure}[h!!!!!]
\begin{center}
 \includegraphics*[width=13cm]     {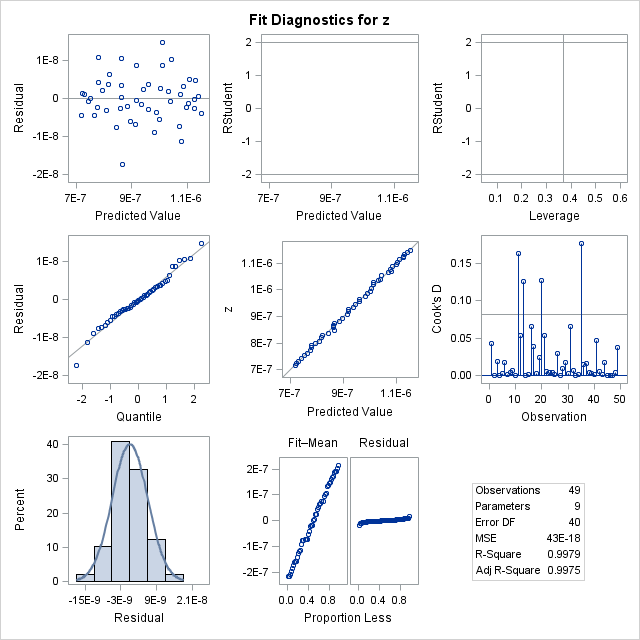}
\caption{Residual plots and other fit diagnostics for the dependent variable (from the SAS output).}
\label{res}
\end{center} 
\end{figure}

Based on the standard diagnostics provided by the SAS regression procedure, we can quantify the reliability and accuracy of the model \eqref{mex}. Specifically, the normalized predicted value for residuals (first panel in Figure~\ref{res}), residual quantile plot, and the predicted value for the dependent variable, all indicate that the model is accurately describing the total variability in the data, and that the canonical normality assumptions are satisfied. 

To further assess the robustness of \eqref{mex}, we employed multiple cross-validation by partitioning the full data set into smaller, equal-sized sets, and then fitting  the model using all but one of the subsets. The predicted values for the missing observations (from the subset removed from the analysis) are then computed and quantified using their mean value and dispersion. 

Specifically, we divided the data set into 49 data sets and use all 48 sets to
construct the model and validate the model using the one left out. Then we repeat the
procedure 48 times. The mean of the residuals is 5.427845e-22 and the variance of the residuals is 2.806e-44.
%
%\begin{figure}[h!!!!!]
%\begin{center}
%% \includegraphics*[width=9.5cm]     {co_circulation.jpg}
% \includegraphics*[width=12cm]     {ResidualPlot.png}
%\includegraphics*[width=12cm]     {ResidualPlot1.png}
%% \includegraphics*[width=6.25cm]     {qq_co2.jpg}
%\caption{Residual plots and other diagnostics for the regressors of the dependent variable (from the SAS output).}
%\label{res1}
%\end{center} 
%\end{figure}

\section{Comparing the US and EU models}

Using the results obtained in \cite{article1}, it is possible develop a comparison between the US and EU quadratic models for attributable variables and interactions; in particular, it is possible to compare the relative relevance of the main single-factor variables and of the main interactions (see Table~\ref{comp} below and Table 3.4 from \cite{model}). 

 \begin{center}
 \begin{table}[h!]
  \caption{Comparison of statistical relevance for attributable variables and interactions, US vs. EU.}
  \label{comp}
 \begin{tabular}{| c || c | c |}
 \hline 
${\mbox{Rank}}$ & ${\mbox{Variable in US}} $  & ${\mbox{Variable in EU}} $  \\
 \hline 
\hline 
$  1 $  & Liquid  & Gas \\
\hline 
$  2 $  & Liquid:Cement & Gas:Bunker  \\
\hline 
$  3 $  & Cement:Bunker  & Liquid:Liquid \\
\hline 
$  4 $  & Bunker  & Bunker:Bunker \\
\hline 
$  5 $  & Cement   &  Gas Flares \\
\hline 
$  6 $  & Gas Flares   & Liquid:Gas Flares  \\
\hline 
$  7 $  & Gas  & Liquid:Bunker \\
 \hline 
$  8 $  & Gas:Gas Flares &  Liquid \\
 \hline 
 \end{tabular}
 \end{table}
\end{center}

%Based on these measures, we can draw a number of immediate conclusions:

%\begin{itemize}
%\item[i)] 

\subsection{Specific features of CO$_2$ contributors for EU data}

As Table~\ref{comp} indicates, the most significant risk factor found when studying the US data (Liquid fuels) is the least relevant in the case of the EU data; likewise, the main attributable variable obtained for the EU data (Gas fuels) is the least relevant single-factor in the case of the US data (rank 7 out of 8). In view of the importance of the variable Gas fuel for the EU data, we indicate the individual contributions at country-level, for the year 2008, in Table~4.

This important difference between the rankings of risk factors for EU and US suggests two directions for further comparisons: 

\begin{itemize}
\item[i)] for the purpose of CO$_2$ emission reductions, a very different picture emerges from the EU study versus the US study. Consider the information shown on Tables 2, 3, and 4: almost 50\% of the total variability in the atmospheric CO$_2$ is due to Gas Fuels alone, which ranks first among the risk factors; furthermore, more than 25\% of the total emissions for this factor is due to a single country, Germany (for year 2008). %Therefore, it appears obvious that significant changes in the overall EU emissions could be achieved by implementing policy changes at single-country levels. 
\item[ii)] when developing criteria for trade-off of single risk factors (as in ``cap-and-trade" schemes and beyond), specific information from each continent must be employed, as discussed in the next section. While this point further emphasizes the regional aspect of CO$_2$ analysis, it also illustrates the inherent limitations of applying carbon-accountability policies from one continent to another. 
\end{itemize}

%\newpage

 \begin{center}
 \begin{table}[h!] \label{t2}
  \caption{Contributions by country to gas fuel emissions, for year 2008 (in thousand metric tons of carbon).}
 \begin{tabular}{| l || l | c | c |}
 \hline
Rank & Country & Total & Percent  \\
 \hline
 \hline 
 1 &	GERMANY	& 85457 & 27.19\%  \\ \hline
2	& POLAND	& 58395 & 18.58\%  \\ \hline
3	& UK 	& 37306 & 11.87\%  \\ \hline 
4	& CZECH REP.	& 21021 & 6.69\%  \\ \hline
5	& ITALY	& 16855 & 5.36\%  \\ \hline
6	& SPAIN	& 14669 & 4.67\%  \\ \hline
7	& FRANCE	& 13383 & 4.26\%  \\ \hline
8	& ROMANIA	& 9805 & 3.12\%  \\ \hline
9	& GREECE	& 8946 & 2.85\%  \\ \hline
10	& BULGARIA	& 8058 & 2.56\%  \\ \hline
11	& NETHERLANDS	& 7346 & 2.34\%  \\ \hline
12	& FINLAND	& 5478 & 1.74\%  \\ \hline
13	& BELGIUM	& 4361 & 1.39\%  \\ \hline
14	& SLOVAKIA	& 4205 & 1.34\%  \\ \hline
15	& DENMARK	& 4062 & 1.29\%  \\ \hline
16	& AUSTRIA	& 3896 & 1.24\%  \\ \hline
17	& HUNGARY	& 3275 & 1.04\%  \\ \hline
18	& PORTUGAL	& 2644 & 0.84\%  \\ \hline
19	& SWEDEN	& 2530 & 0.80\%  \\ \hline
20	& IRELAND	& 2519 & 0.80\%  \\ \hline
21	& LUXEMBOURG 	& 86 & 0.03\%  \\ \hline
22	& CYPRUS	& 29 & 0.01\%  \\ \hline
23	& MALTA	& 0 & 0.00\%  \\ \hline
 % \hline 
 \end{tabular}
 \end{table}
\end{center}

%\item[ii)] 

\subsection{Regionalization of atmospheric CO$_2$ analysis}

One relevant variable for the US market (Cement, together with its interactions to Liquid and Bunker) is completely absent in the case of the EU data. Together with the fact that, at the same level of statistical significance, the US model requires five variables while the EU data leads to a model with four, this shows that the two models are fundamentally different, in the sense that production, dynamics, and global interactions of the man-made factors responsible for atmospheric CO$_2$, are essentially different in the case of EU and UE. 

This indicates that, when developing specific guidelines for industry regulations and accountability criteria, translating regulations and policies between the EU and US markets must be done with considerable caution. The differences found here mandate, in fact, that a specific approach must be developed in the EU case, and that  adopting US-based policy directly may be unwarranted and outright misguided. We will quantify this aspect of the comparison in a forthcoming publication \cite{article3}, aimed at comparing the possible carbon-production management in the case of the US versus the EU markets.

%\item[iii)] Last but not least important, ``trading values" for units of attributable variables cannot be compared directly for these two markets: the US study \cite{article1} led to the conclusion that 1000 units of Gas Flares can be equated to 2127 units of Cement, while in the case of EU, one unit of Liquid is equivalent to approximately 98 units of Gas Flares. However, a direct comparison of the various trading values cannot be established from these models. 
%\end{itemize}

%\section{Conclusions and further studies} 

%\section{Appendix} \label{sec:intro} %\markright{Appendix}
\end{document}